\documentclass[12pt]{article}
\usepackage{sbc-template}
\usepackage{listings}
\usepackage{graphicx,url}
\usepackage[utf8]{inputenc}
\usepackage{caption}

\newcommand{\code}[1]{\texttt{#1}}

\title{Agent Programming for Industrial Applications:\\Some Advantages and Drawbacks\thanks{Supported by Petrobras project AG-BR, IFSC and UFSC.}}

\author{Ot{\'a}vio A. Matoso\inst{1}, Luis P. A. Lampert\inst{1}, Jomi F. Hübner\inst{1},\\Mateus Conceição\inst{1}, S{\'e}rgio P. Bernardes\inst{1}, Cleber J. Amaral\inst{1,2},\\Maicon R. Zatelli\inst{1}, Marcelo L. de Lima\inst{3}}

\address{Universidade Federal de Santa Catarina (UFSC)\\
  Florian{\'o}polis -- SC -- Brazil
\nextinstitute
  Instituto Federal de Santa Catarina (IFSC)\\
  S{\~a}o Jos{\'e} -- SC -- Brazil
\nextinstitute
  Petrobras-Cenpes\\
  Ilha do Fund{\~a}o, Rio de Janeiro -- RJ -- Brazil\\
  \email{\vbox{\noindent otaviomatoso@yahoo.com.br, \{lp.lampert,sergiopb1998\}@gmail.com, \{maicon.zatelli,jomi.hubner\}@ufsc.br, mateusconceicao1@hotmail.com, cleber.amaral@ifsc.edu.br, marceloll@petrobras.com.br}}
}

\begin{document} 

\maketitle

\begin{abstract}
Autonomous agents are seen as a prominent technology to be applied in industrial scenarios. Classical automation solutions are struggling with challenges related to high dynamism, prompt actuation, heterogeneous entities, including humans, and decentralised decision-making. Besides promoting concepts, languages, and tools to face such challenges, agents must also provide high reliability. To assess how appropriate and mature are agents for industrial applications, we have investigated its application in two scenarios of the gas and oil industry. This paper presents the development of systems and the initial results highlighting the advantages and drawbacks of the agents approach when compared with the existing automation solutions.
\end{abstract}

\section{Introduction}

Agents programming offers useful tools to be applied in complex scenarios, with high dynamism and decentralised decision-making, which are common challenges in industrial applications. The agents may take advantage of virtualised information, bringing reactiveness, autonomy and often intelligence in decision-making processes acting over the real devices through their digital representations. In fact, many studies are suggesting the use of Multi-Agent Systems (MAS) in industrial applications~\cite{Dias2017, Kotak2003, Barbosa2018, Mualla2018, Roloff2017, Wada1998, Zhou2007}.


Although the use of agents in industrial scenarios is reasonable, it is not clear whether they can bring more benefits in practical scenarios, especially when compared to classical solutions. This paper presents the development of MAS based solutions for two cases in the gas and oil industry to assess if agents can be applied and which advantages and drawbacks they can bring. The first case regards to a scenario in which a controller must actuate to protect a heat exchanger. The second case regards to adjustments in parameters of the processes that control artificial oil lifting, which are done through the integration of several systems. Both cases are real scenarios and refer to critical processes which are currently automated by classical control solutions.
We have developed prototypes using the JaCaMo\footnote{JaCaMo is an open-source Agent-Oriented Programming platform available in \url{http://jacamo.sourceforge.net}. More details are presented in \cite{boissier2013,boissier:16,boissier19}.} framework for both investigations. In the performed tests the agents use real data usually interacting with real services of the plant.

This paper is organised as follows: Section~\ref{MAS-industry} presents an overview of how MAS are currently being applied in industrial scenarios, Section~\ref{MAS-heatexchange} presents the heat exchangers scenario, the current solution and our agents solution as well as the results obtained from tests. Section~\ref{MAS-lifting} presents the artificial lifting scenario, a list of requirements for an enhanced solution, our MAS solution and results. Finally, Section~\ref{conclusion} summarises our conclusions about the advantages and drawbacks that agents' application may bring for the industry.

\section{MAS in the industry}\label{MAS-industry}

It is often said that the upcoming fourth industrial revolution would be the result of the application of technologies such as Internet of Things (IoT), Internet of Services (IoS), Big Data, Machine Learning, among others~\cite{Rojko2017,klaus_dieter_thoben_2017_1002731, Cimini2017, Lu2017a}. Although these technologies may enhance production, control and relationship with customers in many ways, integrating and managing them all together is a challenge~\cite{Leitao2013, MONOSTORI2016621}.

A factory has many sensors, actuators, controllers, software and other entities that can be virtualised in a standard way, which can allow them to interact with each other.
Many approaches are based on the concepts of Cyber-Physical Systems (CPS). This is a standard abstraction of entities that aims to manage and integrate information and systems, allowing a broad use of data and high integration among all sort of devices.



The agent technology has been successfully applied in a variety of areas in the industry context, namely production planning, scheduling, and logistics.
The agents in such scenarios are bringing reactiveness, autonomy and intelligence. They take advantage of virtualised information for their decision-making processes, acting over real devices through their digital representations.
Multi-Agent Systems (MAS) can be the core of the CPS virtualised entities, providing interoperability and control~\cite{Cruz2017, Leitao2013, Monostori2006, Pechoucek2008}.
It brings several advantages in terms of reconfigurability, scalability, and productivity, which are translated into measurable competitive advantages such as less production time and better use of resources~\cite{Leitao2013}. 
When applied to realise CPS, agent-based approaches leverage their negotiation and communication mechanisms to present benefits regarding flexibility, adaptability, and pro-activity~\cite{Amaral2019b,MONOSTORI2016621,Cruz2017}.

On the other hand, many other issues are still open challenges. There is a lack of suitable, mature simulation and engineering tools, security, safety, real-time capabilities, and standardisation~\cite{Pechoucek2008,Leitao2013}. In addition to the technological barrier, there is also a human factor-related barrier, such as the radical control paradigm shift from a centralised to a modularised control view, with distributed decision-making entities, and the absence of interaction mechanisms to support humans, especially using mobile devices~\cite{Leitao2013,Cruz2017}. 
As a consequence, industrial applications of agents are very limited and their possible benefits in practical means are not well experimented and proven.

\section{Applying agents to protect heat exchangers}\label{MAS-heatexchange}

The first application faced in this work refers to a protection control for heat exchangers used in Floating Production Storage and Offloading (FPSO) processes for the Oil and Gas industry. Heat exchangers are responsible for transferring thermal energy between a fluid and a material which can also be a fluid~\cite{Shah2003}. In FPSO processes, they are used to cool the extracted fluid. Without proper control of the refrigeration, the heat exchange can be exposed to extremes of temperature reducing its lifetime.

\begin{figure}[tp]
	\centering
	\includegraphics[width=0.8\linewidth]{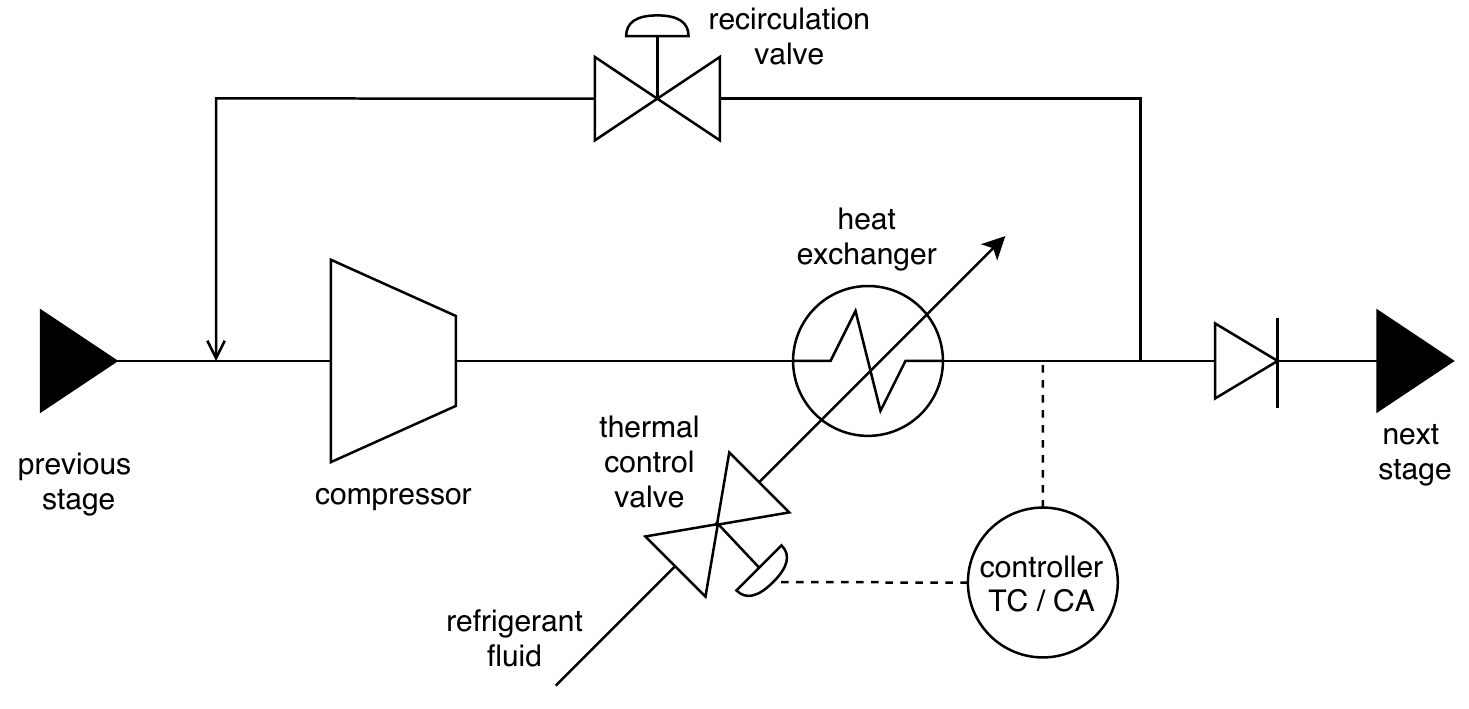}
	\caption{Typical compression stage in FPSO.}
	\label{fig_compressor}
\end{figure}

In the study case (see Figure~\ref{fig_compressor}), the control is performed by two different devices: (i) a Temperature Controller (TC) which is responsible for the processes most of the time controlling the flow of the refrigerant fluid, and (ii) a control application (CA) triggered by special conditions such as compressor starting and stopping processes, unstable and abnormal thermal conditions. The referred situations are usually transients, and the TCs are less effective for them. In this sense, the CA protects the heat exchange until the system reaches a permanent regime and then TC assumes the control. Agents are being used to implement CAs. 

The CA monitors the temperature and assumes the control of the thermal valve, progressively reducing its flows until a previously established set point. The way the valve is controlled varies according to the respective transient nature. The stabilisation process follows a fuzzy logic controller. Orders for changing the valve flow are sent to a system that validates the compensation, logs each order and effectively apply them on the Programmable Logic Controller (PLC) that acts over the physical valve. 

The current CA is developed using classical control devices such as PLCs and supervisory control and data acquisition (SCADA) systems. The logic is implemented using Sequential Flow Charts (SFC)\footnote{Sequential Flow Charts are standardised in the IEC 61131.} which acts over a SCADA or directly over a PLC. The code is modularised into diagrams that are triggered according to preconditions which drive the system to execute the appropriated sequence function. 

\begin{figure}[bph]
	\centering
	\includegraphics[width=0.9\linewidth]{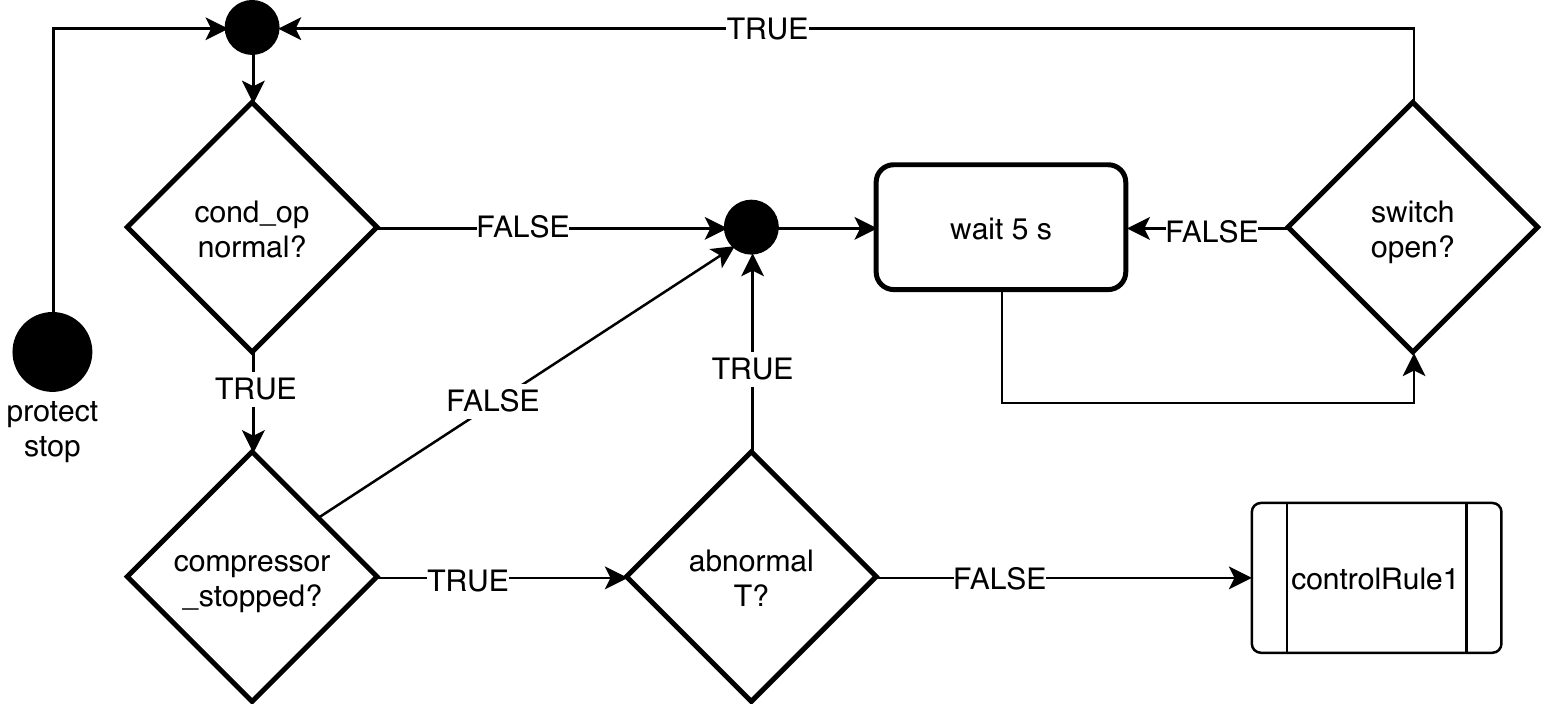}  
	\caption{Current implementation of part of stopping compression process.}
	\label{fig:fig_pollingMPA}
\end{figure}

Figure~\ref{fig:fig_pollingMPA} shows a part of a logic control that protects a heat exchanger when the compression process is going to stop. In summary, the application must take control of the valve when the compressor stops. However, there is a switch that can deactivate the application and a variable that indicates if the system is under operation. Finally, a sensor that indicates an abnormal temperature condition.

The classical logic verifies the state of the variables every 5 seconds. The effective control is performed when it is detected that the compressor has stopped, the controller is on, the system is under an operation condition, and the temperature is normal. In this example, we can verify the use of a polling method to check the condition of certain variables. It is also remarkable the presence of the verification \code{abnormal T?} which regards to a higher priority condition referring to abnormal temperatures. Along with many diagrams of different processes, we can find the same verification since this condition should trigger an overriding process.

In our study, there is no requirement of any enhancement regarding the current control features. The investigation aims to check whether agents can be applied and uncover possible advantages and drawbacks. The cooperation among agents, possibly in a chain of heater exchangers, is out of our scope.

In the solution using agents, we have developed an agent in which most of the CA actuation is programmed. In our case, the agent is coded in Jason, the agent programming framework embedded in JaCaMo. Most of the original implementation have straight-forward counterparts in Jason. However, in some cases, we could take advantage of the Agent-Oriented Programming (AOP) language.

\begin{figure}[tp]
	\centering
	\includegraphics[width=0.5\linewidth]{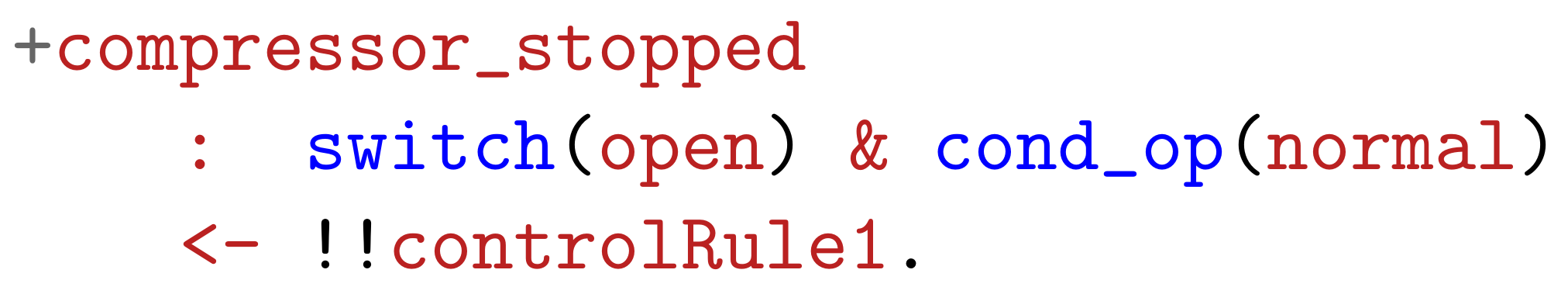}  
	\caption{Agents implementation of part of stopping compression process.}
	\label{fig:fig_pollingMAS}
\end{figure}

Agents are reactive, i.e., they are sensible to events that can trigger actions when perceiving some change in the environment. 
This feature is being exploited in order to avoid polling loops. 
The code in the Figure~\ref{fig:fig_pollingMAS} is an example of how such control logic can be translated to an AOP language. The code defines a plan in which the event \code{compressor\_stopped} may trigger the execution of \code{controlRule1}. To effectively perform the plan, two conditions must be true: \code{switch(open)} and \code{cond\_op(normal)}. In this sense, the approaches presented in Figure~\ref{fig:fig_pollingMPA} and Figure~\ref{fig:fig_pollingMAS} produce similar effects. Besides the agent implementation being arguably shorter and simpler, it is often faster since agents reacts promptly to events, while the SCF may delay up to 5 seconds.
    
In another aspect, the SFC also has a flag to check if the temperature is abnormal. This is a high priority condition that is present in many other parts of the control logic that is being omitted here. Besides redundancy, the need of such check on many parts of the system increases the risk of design mistakes. In agents solution, there is no need to check on each plan this condition since the agent version is taking advantage of agents' goal-oriented behaviour. The agent is coded with the primary goal of protecting the heat exchanger. In this sense, an abnormal temperature condition is triggering another plan that drops any other intention that the agent may have. Doing so, above all, the agent prioritises to perform the reaction to the referred abnormal condition. In this case, an agent-based solution is safer.

Control logic also makes frequent use of timers. Some of them are used to avoid too much stress on checking a condition as presented in Figure~\ref{fig:fig_pollingMPA}. Other timers are placed to give some timeout for an operation. Other ones are used for supporting some process that needs to filter outlier conditions, and so on. In the case of agents, the timers were also sensibly simplified since Jason supports scheduled operations such as \code{.at(t,e)} which executes \code{e} after \code{t} period. 

With the CA implemented in Jason, we observed a reduction in the code size. An evaluation comparing the CA written on the basis of a literal translation of the classic control logic and the CA written exploiting Jason features has shown that the latter is about 40\% smaller~\footnote{For this evaluation, we have compared the code size after compressing the versions.}. Besides facilities like the scheduled operations, some code has been saved because of several verifications become useless when applying the concept of goals priority. 

Agents also facilitate to enhance system robustness. We have added on the agent some plans for possible, even unlikely, situations. Indeed, agents can have several plans in order to fulfil a single goal, i.e., in a range of possible plans, the agent chooses to perform the first applicable. It means that for many reasons, plans cannot be executable, or they can fail. Still, in a way, one of them can be successful, resulting in the achievement of the goal.

Although AOP helped in the referred situations, we could not find advantages on the agent's version when acting over control variables and when calculating parameters. The SFC solution in very convenient on procedural tasks, including real-time ones. Industry standard solutions also have lots of tools helping to monitor variables of the industrial plant. The programming interface of the classical solution is more friendly as well.

\section{Applying MAS to control artificial lifting}\label{MAS-lifting}

Some oil wells provide enough pressure for the natural lifting of the oil in production flow needs. For wells without enough pressure, especially in deep waters, it is necessary to apply artificial lifting techniques which are usually done by centrifugal pumps or by gas injection. These different artificial approaches are roughly similarly parameterised according to the reservoir pressure condition.

The procedure for changing the actuation of the artificial lifting process follows five main stages: (i) results from new tests on the reservoir are available, (ii) a new mathematical model of the reservoir is drawn, (iii) new control parameters for controlling the flow are defined and optimised, (iv) the control system is set up with new parameters, and (v) an external governmental agency is informed. Due to several uncertainties, the final decision-making process depends on human specialists validations.

\begin{figure}[btph]
	\centering
	\includegraphics[width=0.9\linewidth]{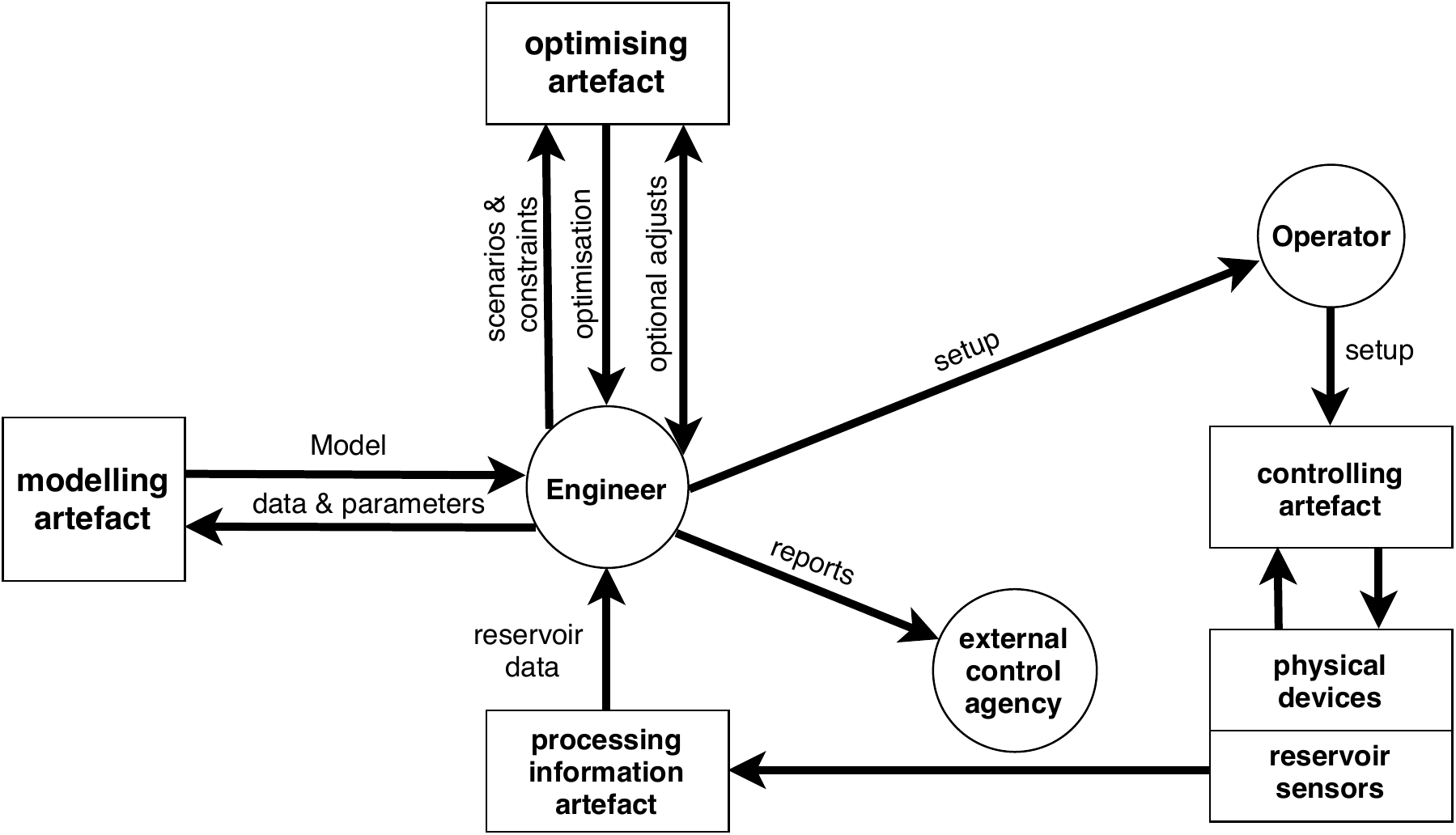}  
	\caption{Classic solution for artificial lifting.}
	\label{fig:liftingClassic}
\end{figure}

Figure~\ref{fig:liftingClassic} illustrates the current process in which the engineer participates on the whole cycle. S/He is responsible for obtaining data, triggering the model design, sending parameters for optimisation and for forwarding setting up commands to the operator. The operator validates the setup with colleagues and then set it up in the control system. The engineer is also responsible for sending to an external agency reports about the production.

An enhanced solution has as main requirement the reduction of the engineer's workload. Although it is mandatory to have a validation of the optimisation from the perspective of an engineer, the coordination tasks can be delegated to an automated system.

Agents' proactivity and goal-oriented behaviour may help with tasks that are not easily automated due to the necessity of interactions and adjustments. Proactive agents chase their goals, in this case, performing the optimisation of the artificial lifting process. Properly developed goal-oriented agents have multiple plans to deal with a sort of circumstances, providing necessary adjusts or starting the process from the scratch if needed.

\begin{figure}[btph]
	\centering
	\includegraphics[width=0.9\linewidth]{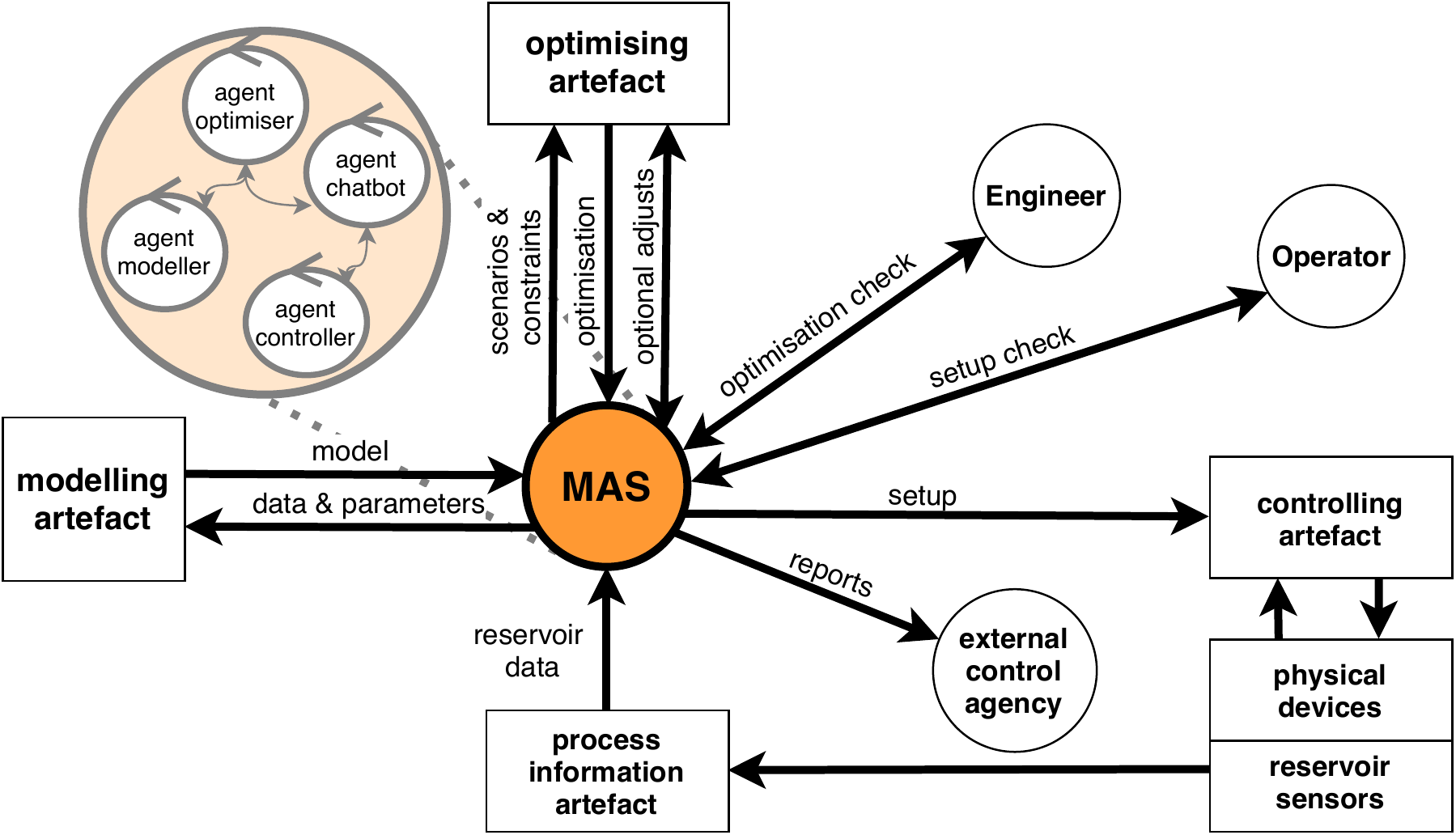}  
	\caption{MAS solution for artificial lifting.}
	\label{fig:liftingMAS}
\end{figure}

Our MAS approach to this problem is presented in Figure~\ref{fig:liftingMAS}. In this solution, the MAS is a new entity in the diagram. It is performing coordination tasks releasing the engineer from this workload. To do so, the MAS is equipped with a mediation tool that helps the agents to communicate with different entities. Some entities are software that perform procedural tasks, i.e., industrial artefacts. Other entities are agents, both human and artificial, that require different communication skills and argumentation capabilities. The system is composed of four agents, as represented in the zoomed area on the top left. These agents play roles in an organisation.

From the communication perspective, the integration between the MAS and external entities is done thought Apache Camel mediation tool. The process information system, the modeller, the optimiser, and the control system are integrated through REST and/or CORBA interfaces. The integration among artificial agents is internally resolved by JaCaMo default infrastructure. The interaction among artificial agents and humans is performed through a chat messaging application which is also integrated with the MAS using Apache Camel. To enhance this interaction, the agent \emph{chatbot} has natural language processing skills. 

Although integration facilities is provided by an agnostic mediation tool, it is important to mention that the MAS solution has the advantage of providing entities model abstractions for both autonomous agents (including humans) and industrial artefacts. 
From the agent’s perspective, it means that it interacts with these entities in a usual way, i.e., using its messaging system if the entity is modelled as an agent or through action and perception if it is modelled as an accessible tool of its environment\footnote{More details of the advantages of such abstractions are available in \cite{Amaral2019} publication.}. 

\begin{figure}[tbp]
	\centering
	\includegraphics[width=1.0\linewidth]{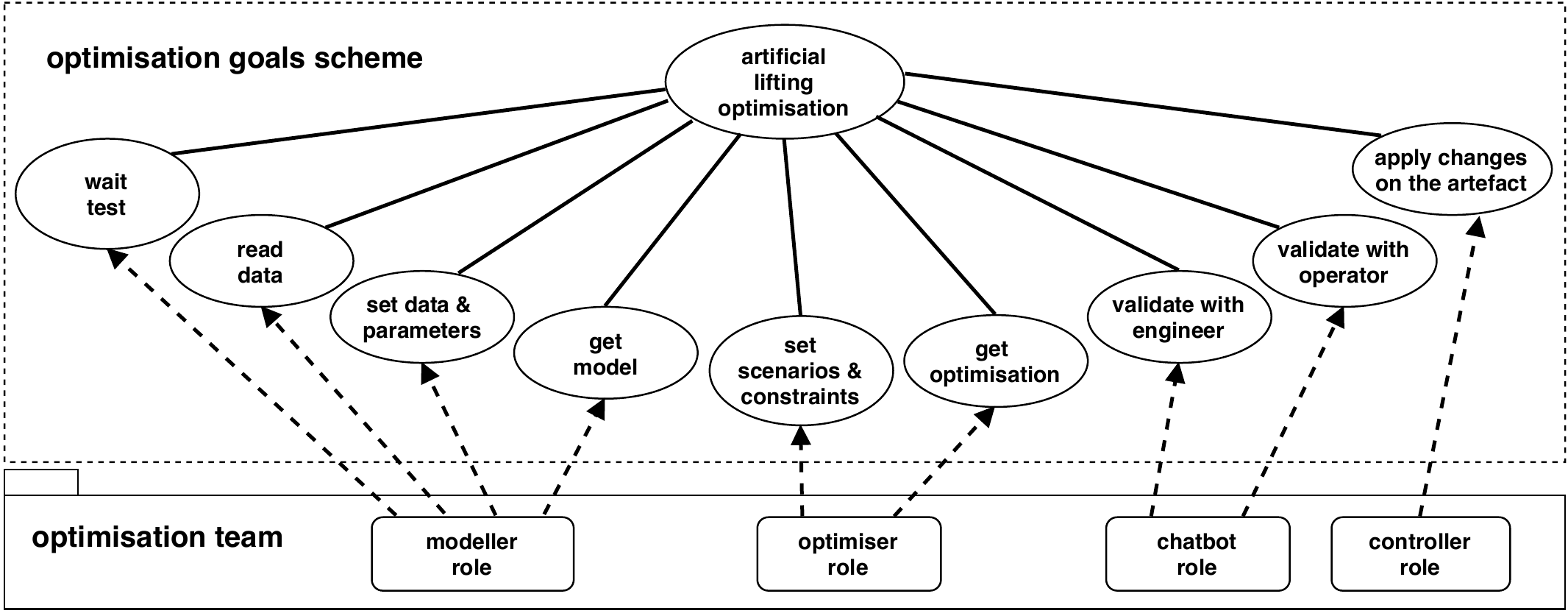}  
	\caption{MAS organisation for artificial lifting.}
	\label{fig:liftingORG}
\end{figure}

In this application, as shown in the Figure~\ref{fig:liftingORG}, agents playing organisational roles are committed to sub-goals of the main goal of optimising the artificial lifting process. The agent \emph{modeller} is in charge of reading reservoir data, sending the data to the modelling artefact for delivering the new model to the agent \emph{optimiser}. This agent sends to the optimisation artefact scenarios obtained from the model and constraints. The artefact answers with an optimisation configuration that is sent to the engineer, through agent \emph{chatbot}. The engineer may contest the parameter which may suffer adjustments. If accepted the agent \emph{chatbot} checks with the operator the proposed setup to be applied. Whether the operator agrees with that, the agent \emph{controller} apply the setup in the controlling artefact.

One advantage of AOP solution regards the clearer division of responsibilities. The division of sub-goal among agents work as a system's modularisation which simplifies significantly individual agent's programming processes. Additionally, the use of an organisational apparatus provides coherence among agents' actions. The coordination of agents by such structure can also simplify rearrangements since sub-goals interactions and dependencies are all defined in a higher abstraction. Indeed, the organisation programming dimension promotes development of macro level features which allows to adjust the behaviour of the system without changing the code of agents and artefacts. For instance, if it is requested to ask the engineer a final validation after the operator's answer, there is no need to change agents' code, but adding to the organisational scheme another sub-goal.

\section{Conclusion}\label{conclusion}

In the industrial applications considered in this initial investigation, we can notice some enhancements when agents are applied. In the heat exchangers scenario, the reactive characteristic of agents has shown that the system can respond to environmental changes quickly (no need for polling). However, this is not an exclusive feature of AOP, it is present in any event-driven programming approach. Agent advantages became clearer when it is added some goal-oriented behaviour, which expands the system's capabilities to deal with environment dynamism. Agents can be programmed to be ready for all sorts of circumstances that may occur while they are achieving their goals. Some situations may also change the entire course of the agent's actions as we have illustrated in Section~\ref{MAS-heatexchange} when current intentions were dropped to achieve a higher priority goal. In this sense, AOP version has shown it can be faster and safer.

It is difficult to measure how easy it is coding in a programming language. However, at least taking the comparison in terms of code size, we realise that the code of an agent application can be smaller than an equivalent application coded in a classic control approach. Although, AOP program can be shorter and simpler, when considering legibility, the code reduction is competing against a visual solution that uses SFC diagrams. A visual programming interface can steep the learning curve and it is often preferable for designers. It is also to be mentioned that constructors used in those diagrams are common tools for most engineers, while predicate logic is not so usual.

The second case brought us other situation in which it was expected to enhance the system in order to reduce the workload over the engineer. This case regards to a complex information flow in a chain of processes that requires coordination and decision-making capabilities from a central actor. We have proposed a MAS in this position since, although complex, the requirements follow standards that can be delegated to agents. The first advantage we have noticed regards to the abstractions for autonomous and non-autonomous external entities AOP is providing. The acknowledgement of the nature of the external entity may boosts the integration through proper communication methods. It can also enhance decision-making processes, for instance, with a better understanding of actions determinism. Another advantage regards to the use of a set of agents as part of an organisation which helps to divide the complexity and allows faster changes in the system, for instance, in the sequence of the actions. In the case of needing a double-check on the setup parameterisation, we have shown that with AOP it is only necessary to introduce this new stage in the goals scheme. No changes in the agents would be needed. 
It seems to be an appropriate facility since the introduction of a double-check is closer to a coordination functionality, than to an individual agent's business rule. 
In this sense, besides an arguably better abstraction, this coordination apparatus remarkably improves flexibility.

Besides some advantages, we experienced difficulties in debugging the MAS approach when integrated directly to control devices. Indeed, the MAS has no tools for such supervision and presents no advantages in executing procedural tasks. Although in a fair comparison between paradigms we should not consider the maturity level of their tools, AOP did not bring advantages in procedural tasks. In this case, in a pragmatic point of view, the lack of tools and the cost of change can be considered AOP drawbacks.

An important issue one may be concerned about the MAS adoption is related to its reliability. Classic control systems are highly reliable. In MAS, it is possible to develop agents to monitor other agents in order to replace them in case of fails, for instance. However, in this investigation, unfortunately, we could not test such a possibility and did not stress the MAS or use it broadly, for instance, in several heat exchangers or in long-term plant operation. In this sense, we still have no evidence that MAS is reliable and robust enough. 

In this sense, we find the MAS solution a prominent approach for the core of the CPS. Acknowledging the advantages of classical solution and the need of more evidences on reliability of agents, we find a better solution by combining both approaches (agents and classical) together. We opt to give the agents the responsibility for events, timers, and for triggering smaller and simplified versions of SFCs. They are also responsible for coordination processes. The control procedures, including real-time ones, remain on classical devices.

As future work we intend to: (i) exploit advantages on cooperation among agents in a chain of heat exchangers, (ii) adapt the agent platform to be integrated with monitoring tools, (iii) develop fault tolerance facilities such as agents' health monitoring and actuators, (iv) develop a unit testing tool for agents, and (v) expand system tests for long-term running sandboxes and real plants for checking issues such as robustness and reliability.

\bibliographystyle{sbc}
\bibliography{sbc-template}

\end{document}